\documentclass[conference]{IEEEtran}
\IEEEoverridecommandlockouts
\usepackage{cite}
\usepackage{amsmath,amssymb,amsfonts}
\usepackage{graphicx}
\usepackage{textcomp}
\usepackage{xcolor}
\usepackage{tabularx}
\usepackage{booktabs}

\usepackage{ragged2e} 
\usepackage{makecell}
\usepackage{multirow}

\usepackage{array}

\usepackage{verbatim}
\usepackage{makecell}
\usepackage{graphicx}
\usepackage{algorithm}
\usepackage{algpseudocode}
\usepackage[caption=false]{subfig}
\usepackage{float}

\newfloat{figtab}{htb}{fgtb}
\makeatletter
  \newcommand\figcaption{\def\@captype{figure}\caption}
  \newcommand\tabcaption{\def\@captype{table}\caption}
\makeatother

\def\BibTeX{{\rm B\kern-.05em{\sc i\kern-.025em b}\kern-.08em
    T\kern-.1667em\lower.7ex\hbox{E}\kern-.125emX}}

\begin{document}

\title{Energy-Efficient WiFi Backscatter Communication for Green IoTs}

\author{
 \IEEEauthorblockN{Yimeng Huang\IEEEauthorrefmark{1}\IEEEauthorrefmark{2}, Lijie Liu\IEEEauthorrefmark{1}, Jihong Yu\IEEEauthorrefmark{3}, Yuguang Fang\IEEEauthorrefmark{2}, Wei Gong\IEEEauthorrefmark{1}\IEEEauthorrefmark{4}}
 \IEEEauthorblockA{\IEEEauthorrefmark{1}University of Science and Technology of China, China\\
 \IEEEauthorrefmark{2}City University of Hong Kong, China\\
 \IEEEauthorrefmark{3}Beijing Institute of Technology, China\\
  huangyimeng@mail.ustc.edu.cn, lijieliu@mail.ustc.edu.cn, jihong.yu@bit.edu.cn, \\ my.fang@cityu.edu.hk,
   weigong@ustc.edu.cn}
  \thanks{\IEEEauthorrefmark{4}Corresponding author: Wei Gong.}
 }

\maketitle

\begin{abstract}
%The boom in the Internet of Things (IoT) has greatly improved the quality of our life but it is also accompanied by massive consumption of resources and pollution.
%%This issue forces us to look for new design concepts to minimize the consumption of IoT systems.
%In recent years, the concept of Green IoT (GIoT) has become a worldwide consensus.
%Backscatter is a promising way to achieve this vision. 
%However, we find that the overall power consumption of existing +systems is still high due to the reliance on redundant transceivers.
%In this paper, we present EEWscatter, an energy-efficient WiFi backscatter that minimizes consumption in every part of the system to meet the requirements of GIoT.
%Specifically, we reuse the existing ambient data and embed tag data in an ultra-low power way by backscatter.
%Further, we devise two new CRC-based methods to demodulate both ambient and tag data on a single receiver.
%Results show that our system saves over half the consumption of the systems in place.
%We also demonstrate that the system time complexity can be as low as O(n) consuming just a little computing resource.
%%For transmission, the BER achieves $\textbf{10}^{\textbf{-2}}$ when SNR $>$ 30.
%Based on green design principles, we believe EEWscatter is a vital step towards a green world.

%0414 chat
The boom of the Internet of Things has revolutionized people's lives, but it has also resulted in massive resource consumption and environmental pollution. 
Recently, Green IoT (GIoT) has become a worldwide consensus to address this issue.
In this paper, we propose EEWScatter, an energy-efficient WiFi backscatter communication system to pursue the goal of GIoT.
Unlike previous backscatter systems that solely focus on tags, our approach offers a comprehensive system-wide view on energy conservation.
Specifically, we reuse ambient signals as carriers and utilize an ultra-low-power and battery-free design for tag nodes by backscatter.
Further, we design a new CRC-based algorithm that enables the demodulation of both ambient and tag data by only a single receiver while using ambient carriers.
Such a design eliminates system reliance on redundant transceivers with high power consumption. 
Results demonstrate that EEWScatter achieves the lowest overall system power consumption and saves at least half of the energy.
What's more, the power consumption of our tag is only 1/1000 of that of active radio.
We believe that EEWScatter is a critical step towards a sustainable future. 
%0414 chat

\end{abstract}


\section{Introduction}
%The Internet of Things has facilitated our lives greatly, leading to a globally connected world. 
%Wireless Sensor Networks (WSNs) build a bridge between the digital and physical worlds by deploying a variety of sensors to collect surrounding information.
%The number of connected devices has increased dramatically and is expected to reach 30 billion by 2030.
%%\cite{number-of-IoT-devices}.
%However, the boom of IoT is achieved at the price of huge energy consumption and prodigious greenhouse gas emissions, e.g. carbon dioxide (CO2). 
%The carbon footprint of information and Communication Technology (ICT) accounts for approximately 3\% of the global energy consumption \cite{guegan2019estimating} and 3.7\% of global greenhouse emissions \cite{carbon-footprint-web} in recent years. 
%What's worse, it will increase drastically in the near future if no effective measures are taken.

%0414 chat
The Internet of Things (IoT) has improved our lives greatly, connecting us to a global network. The number of connected devices is surging and is predicted to reach 30 billion by 2030. However, the rapid expansion of IoT comes at a high cost of energy consumption and carbon dioxide (CO2) emissions. The carbon footprint of Information and Communication Technology (ICT) accounts for approximately 3\% of global energy consumption \cite{guegan2019estimating} and 3.7\% of global greenhouse gas emissions \cite{carbon-footprint-web} in recent years. If no effective actions are taken, this figure is expected to rise significantly in the future.
%0414 chat

%Fortunately, this issue has attracted worldwide attention and Green IoT \cite{gandotra2017green}, a new concept, has been introduced to facilitate reducing energy consumption and carbon emission to achieve a sustainable smart world. 
%GIoT refers to exploring cutting-edge ICT to save energy used for designing, manufacturing, and operating IoT equipment \cite{carbon-footprint-web}. 
%Key components we need to focus on of GIoT are shown in Fig.~\ref{giot}, including communication systems, networking architectures, transceiver devices, and so on.
%So what kind of energy-efficient communication is needed for Green IoT and how to achieve it? 
%Backscatter, an advanced communication technique that can transmit data in a battery-free way, is a viable solution \cite{zhang2016hitchhike}  \cite{zhao2018spatial} \cite{zhang2017freerider}. 
%However, the energy efficiency of existing systems is far from ideal as they only focus on the tag but ignore the overall system.
%Actually, the power consumption of other devices like transceivers is non-negligible and much higher than tag.
%Therefore, a system-wide perspective is necessary.
%We take a deep look at the three main parts of a backscatter communication system and try to reach the minimum limit of resource usage for each part.

\begin{figure}[t]
	\centering
	\includegraphics[width=0.55\linewidth]{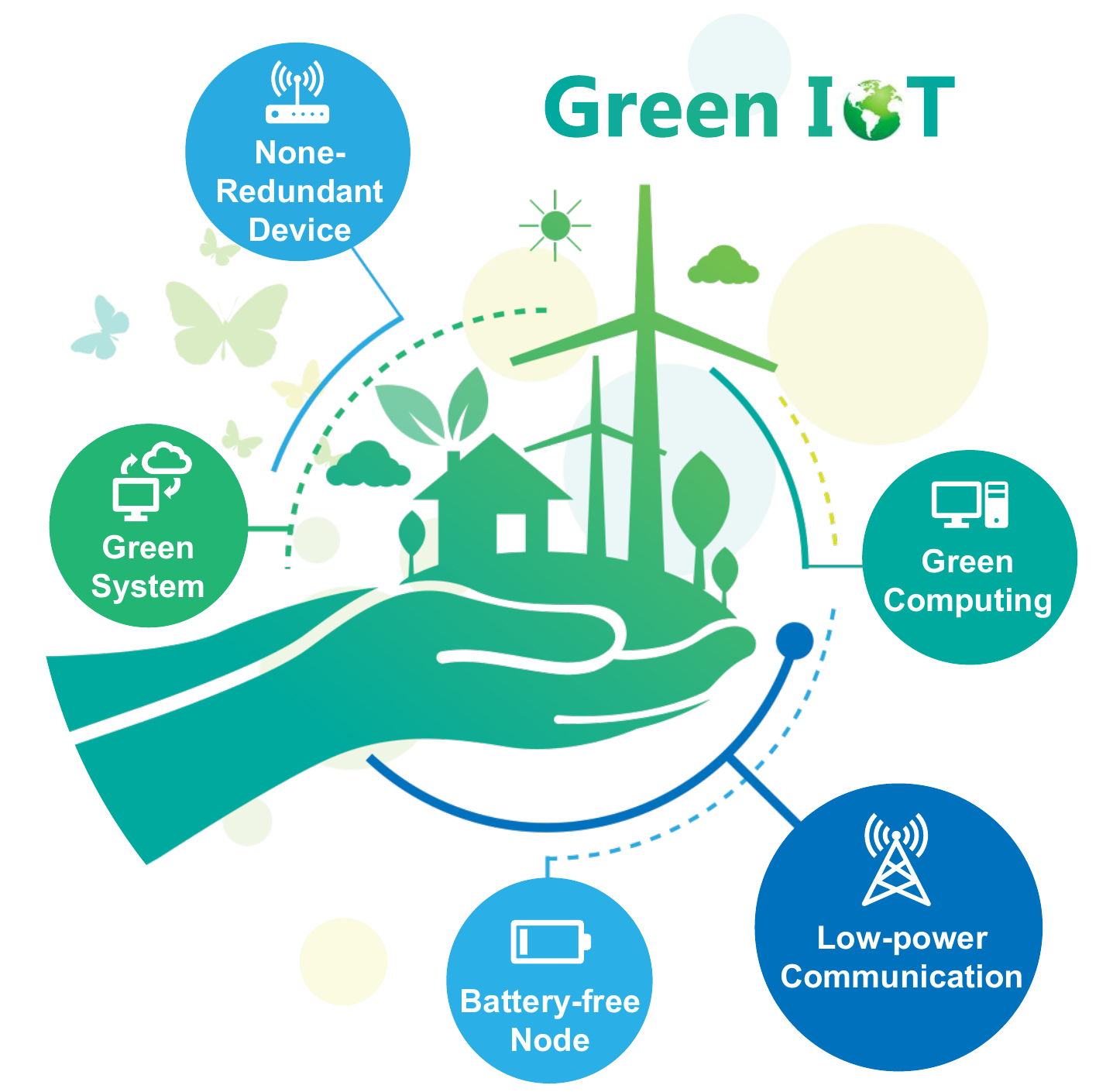}
	\caption{The vision of Green IoT achieved by backscatter.}
	\label{giot}
\end{figure}

%0414 chat
Fortunately, this issue has garnered global attention, and a new concept, Green IoT \cite{gandotra2017green}, has been introduced to promote sustainable smart world development. 
Green IoT is aimed at leveraging state-of-the-art ICT to reduce energy consumption and environmental pollution brought by the IoT industry. 
Backscatter technology \cite{zhang2016hitchhike} \cite{zhao2018spatial} \cite{zhang2017freerider} \cite{kellogg2014wi} is a promising low-power communication solution that can help achieve this goal, as illustrated in Fig.~\ref{giot}.
The core component, backscatter tags, eliminates the need for batteries and reduces the device requirements by utilizing existing RF radios.
The passive design of the tag also enables it to work in an ultra-low-power mode, conserving computing resources.
With the increasing focus on green development, backscatter systems is likely to become the main force for green communication in the future.
However, current backscatter systems primarily focus on the tag, overlooking other components.
Therefore, there are still rooms to be improved and a system-wide perspective is necessary to achieve optimal energy efficiency. 
This paper takes a comprehensive approach by examining the three primary components of a backscatter communication system and aiming to minimize resource usage for each component.
%0414 chat

\paragraph{Energy-efficiency for Excitor}
%In a backscatter system, an excitor is used to provide carrier signals, which are provided in two main ways. 
%Some systems use a transmit helper to generate specific carriers, e.g., single tone. 
%But the additional equipment brings extra overhead.
%A better choice is to take advantage of already deployed devices.
%By reusing the existing ambient signals, the extra cost is almost zero.
%So, we use the ambient OFDM WiFi as excitation to converge on the GIoT goal.

%0414 chat
Backscatter systems require an excitor to provide carrier signals, which can be generated in two ways. Some systems rely on an extra transmit helper to generate specific carriers, such as single tone, so this approach comes with additional equipment and overhead. Alternatively, leveraging deployed devices provides a better solution. By reusing existing signals, the extra cost is almost zero. Therefore, we prefer using the ambient signals as the excitation to reach the GIoT design goal.
%0414 chat

%0414 chat
\paragraph{Energy-efficiency on Tag}
Traditional active wireless terminal devices use high-power-consuming modules for data transmission, so they require additional power supply (battery) to keep the system running. 
It often leads to the massive use and disposal of batteries, causing serious environmental pollution. 
As an alternative, we propose using backscatter tag, It's a battery-free solution that conveys tag data on carriers with ultra-low power.
%0414 chat

\paragraph{Energy-efficiency for Receiver}
%The duty of receivers is to capture and decode the backscattered signal.
%In some systems, two receivers are needed bringing hardware redundancy and additional consumption for signal synchronization.
%By special design, some other systems can decode tag data using only one device.
%But obtaining the tag data alone still can't keep the transmission of original ambient data that needs other communication traffic.
%Hence, we aim to minimize the energy consumption for the receiver by decoding both tag data and ambient data with a single AP.

%0414 chat
In backscatter systems, the receivers are responsible for capturing and decoding the backscattered signal. Some systems require two receivers, which adds hardware redundancy and increases consumption for signal synchronization. Some others can decode tag data using only one device, but cannot sustain the transmission of original ambient data. Our aim is to minimize energy consumption at the receiver by decoding both tag data and ambient data with a single receiver.
%0414 chat

%In this paper, we propose EEWScatter, an energy-efficient WiFi backscatter communication system that achieves minimized power consumption of the whole system. 
%We reuse the existing ambient WiFi signal that brings near-zero additional burden. 
%%modify 0413 
%Battery-free tags with only 271 $\mu$W of power consumption are used to realize low-power transmission. 
%%modify 0413
%We also design two novel CRC-based demodulation methods that enable one receiver to recover both the ambient and tag data. 
%The results demonstrate that our system consumes only 1/2 the power of existing backscatter systems. 
%EEWScatter provides a viable solution for achieving GIoT.

%0414 chat
In this paper, we present EEWScatter, an energy-efficient WiFi backscatter communication system that makes the best use of available resources and takes us closer to the vision of GIoT.
Results show that we save at least half of the power of existing backscatter systems.
The main contributions are summarized as follows:

\begin{enumerate}
    \item We are the first to consider power consumption from a system-wide view and find the essential reason why current systems fail to achieve optimal resource utilization.

    \item We propose a new CRC-based demodulation method that eliminates receiver redundancy, enabling only one receiver to decode both the ambient and tag data. 

    \item We propose a feasible solution that minimizes the power consumption for all three components above, thereby reducing overall system power consumption.
\end{enumerate}

%0414 chat

\begin{figure} [t]
	\centering
	\includegraphics[width=0.97\linewidth]{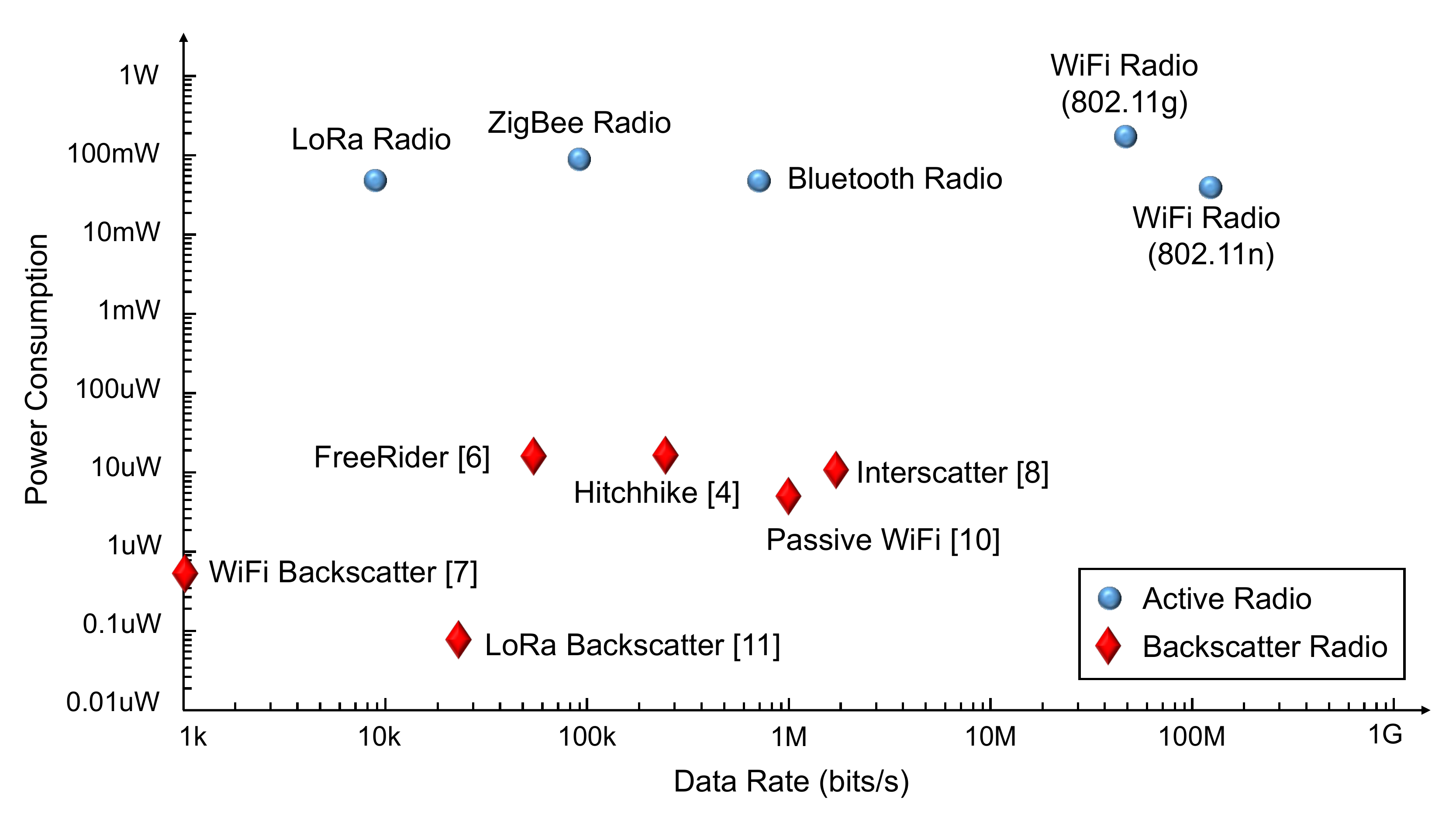}
	\caption{Comparison of tag power consumption.}
	\label{tag_power}
\end{figure}

\begin{comment}
\renewcommand\arraystretch{1.28}
\begin{table}[t]
\caption{Transciver Comparison of Backscatter Systems}
\begin{center}
\begin{tabularx}{\linewidth}{ m{2.7cm}<{\reggedleft}  m{1.5cm}<{\centering}  m{1.5cm}<{\centering}  m{1.5cm}<{\centering}}
\hline
\vspace{5pt}
{} & {Productive Carrier} 
& {Single Receiver}  
& {Uncontrolled Content} \\
\hline
{Passive WiFi \cite{kellogg2016passive}} & {} & {$\bullet$}  & {} \\
{LoRa Backscatter \cite{talla2017lora}} & {} & {$\bullet$}  & {} \\
{BLE-Backscatter \cite{ensworth2017ble}} & {} & {$\bullet$}  & {} \\
{Interscatter \cite{iyer2016inter}} & {} & {$\bullet$}  & {} \\
{Hitchhike \cite{zhang2016hitchhike}} & {$\bullet$} & {}  & {$\bullet$} \\
{MOXcatter \cite{zhao2018spatial}} & {$\bullet$} & {}  & {$\bullet$} \\
{X-Tandem \cite{zhao2018x}} & {$\bullet$} & {}  & {$\bullet$} \\
{PLoRa \cite{peng2018plora}} & {$\bullet$} & {}  & {$\bullet$} \\
%{BackFi \cite{bharadia2015backfi}} & {$\bullet$} & {}  & {$\bullet$} \\
%{RapidRider+ \cite{wang2021rapidrider}} & {$\bullet$} & {$\bullet$}  & {} \\
\hline
\vspace{2pt}
{EEWscatter} & {$\bullet$} & {$\bullet$}  & {$\bullet$} \\
\hline
\end{tabularx}
\label{transceiver-comparison}
\end{center}
\end{table}
\end{comment}

\section{Motivation}
With the dramatic increase in IoT devices, the enormous resource consumption has placed a heavy burden.
To address this issue, we first examine why current backscatter systems consume excessive energy, and then offer a possible green solution to reach GIoT vision.

\subsection{Active Radio Consumes More}
A large number of sensors are widely deployed to gather information and provide data support for intelligent analysis and control.
Most of them transmit data using active radios with high power consumption as shown in Fig.~\ref{tag_power}.
All of them are more than 10 mW due to energy-consuming components like amplifiers, mixers, etc.
So it is necessary to use batteries or other energy sources to power the devices.
It is evident that the disposal of a significant number of batteries, once depleted, could lead to serious environmental hazards.

To address the challenges of high energy consumption and environmental pollution caused by active radios, we propose to use the backscatter technique. 
This technique has been a breakthrough in enabling battery-free devices and achieving near-zero power consumption through the use of passive components such as resistors and capacitors.
To evaluate the power consumption of advanced backscatter systems, we present a comparison in Fig.~\ref{tag_power} with data obtained from several systems, including Hithhike \cite{zhang2016hitchhike}, FreeRider \cite{zhang2017freerider}, and Interscatter \cite{iyer2016inter}.
These systems exhibit power consumption at the microwatt-level. 
We observe that backscatter systems are at least 1000x more energy efficient than corresponding active radios \cite{talla2021advances}.
%\cite{talla2021advances}. 

However, despite backscatter technology significantly reducing the power consumption of the tag, existing systems often fail to consider the power consumption of the excitor and receiver components, which can be much higher than that of the tag. 
This oversight may have a significant impact on the overall power efficiency of the system.
Next, we will undertake further investigations of the transceiver to minimize the overall power consumption of the system.

\begin{comment}
    \begin{table}[t]
    \caption{Transciver Comparison of Backscatter Systems}
    \begin{center}
        \begin{tabularx}{\linewidth}
                { m{2.7cm}<{\reggedleft} 
                  m{2.5cm}<{\centering}  
                  m{2.7cm}<{\centering} }
            \hline
            \vspace{7pt}
            {} & {Excitor Consumption} 
            & {Receiver Consumption}  \\
            \hline
            {Passive WiFi \cite{kellogg2016passive}} & {High} & {Low}   \\
            {LoRa Backscatter \cite{talla2017lora}} & {High} & {Low}   \\
            {Interscatter \cite{iyer2016inter}} & {High} & {Low}   \\
            {Hitchhike \cite{zhang2016hitchhike}} & {Low} & {High}  \\
            {FreeRider \cite{zhang2017freerider}} & {Low}  & {High} \\
            {X-Tandem \cite{zhao2018x}} & {Low} & {High}  \\
            \hline
            \vspace{2pt}
            {EEWScatter} & {Low} & {Low}  \\
            \hline
        \end{tabularx}
    \label{transceiver-comparison}
    \end{center}
\end{table}
\end{comment}
\renewcommand\arraystretch{1.2}
\begin{table}[t]
    \centering
    \caption{Transciver Comparison of Backscatter Systems} \label{transceiver-comparison}
    {
    %\begin{center}
        \begin{tabularx}{\linewidth}{
                m{2.7cm}<{\raggedright}
                m{2.5cm}<{\centering}
                m{2.7cm}<{\centering}
        %\begin{tabularx}{0.45\textwidth} {
         %>{\raggedright\arraybackslash}X
         %>{\centering\arraybackslash}X
         %>{\centering\arraybackslash}X
         %>{\setlength{\linewidth}{1.025\linewidth}\raggedright}X     %>{\setlength{\linewidth}{0.95\linewidth}\centering}X
         %>{\setlength{\linewidth}{1.025\linewidth}\centering}X
        }
            \hline
            %\vspace{7pt}
            {} & {Excitor Consumption} 
            & {Receiver Consumption}  \\
            \hline
            {Passive WiFi \cite{kellogg2016passive}} & {High} & {Low}   \\
            {LoRa Backscatter \cite{talla2017lora}} & {High} & {Low}   \\
            {Interscatter \cite{iyer2016inter}} & {High} & {Low}   \\
            {Hitchhike \cite{zhang2016hitchhike}} & {Low} & {High}  \\
            {FreeRider \cite{zhang2017freerider}} & {Low}  & {High} \\
            {X-Tandem \cite{zhao2018x}} & {Low} & {High}  \\
            \hline
            %\vspace{2pt}
            {EEWScatter} & {Low} & {Low}  \\
            \hline
        \end{tabularx}
    }
    %\end{center}
\end{table}

\subsection{Redundant Transceiver Consumes More} \label{redundant transceiver}

Due to the limitations in tag data modulation and decoding technology, ancillary devices are necessary to facilitate tag data transmission in the current systems leading to unnecessary overhead. 
These systems fall under two primary categories.

\paragraph{Rely on Transmitter Helper}
The first category uses a helper device to generate single-tone (continuous sine wave) as carrier signals, upon which the tag modulates the desired content.
For example, Passive WiFi \cite{kellogg2016passive} and LoRa backscatter \cite{talla2017lora} %and BLE-backscatter \cite{ensworth2017ble} uses a 
dedicated plug-in device to generate single-tones.
Interscatter \cite{iyer2016inter} uses reverse whitening techniques to turn a Bluetooth signal into a partial single-tone.
In this manner, the receiver can recover the tag data directly from the backscattered signal since the carrier does not contain any information.
Therefore, these systems heavily depend on specific excitor, rendering them ineffective when alternative excitation signals are utilized. 
Table \ref{transceiver-comparison} demonstrates that the utilization of a redundant transmitter helper leads to a notable rise in the excitor power consumption, which also elevates the overall power consumption of the system.

\paragraph{Rely on Two Receivers}
The pioneering work, namely Hitchhike \cite{zhang2016hitchhike}, opens the door for getting rid of extra excitor.
\cite{zhang2016hitchhike} proposes codeword translation that enables to reuse ambient 802.11b signals as carriers.
A number of papers have been inspired by this and made further contributions, e.g., MOXcatter \cite{zhao2018spatial}, X-Tandem \cite{zhao2018x}, and PLoRa \cite{peng2018plora}.
%, BackFi \cite{bharadia2015backfi}. 
Yet, codeword translation requires both backscatter data and ambient data to decode tag data.
These systems cannot recover ambient data from backscatter signal, so two receivers are needed to capture the two types of data, separately.
The additional receiver increases the overall power consumption of the system.
%Not only does this lead to an increase in the number of devices, but also the synchronization between two devices brings additional computational overhead.

\begin{comment}
\item{To further reduce device redundancy, some advanced works try to decode tag data by only one receiver.
RapidRider+ \cite{wang2021rapidrider} enables that by adding redundancy to the ambient data so that the receiver can get both ambient data and tag data in one packet.
The extra-control applied to ambient data affects the original communication functionality of the device.
So additional traffic is needed to enable communication increasing the power consumption of the device.}
\end{comment}

%Therefore, the common problem with the current systems is transceiver redundancy.
%Generally, the power of transceivers is almost $10^6$ times higher than the tag.
%Therefore, the redundancy device will greatly increase the cost and power consumption of the communication system.
%To pursue GIoT, we must pay special attention to the number of transceivers in use.

Hence, the prominent issue with contemporary systems pertains to transceiver redundancy. 
In general, transceivers consume nearly $10^6$ times more power than tags, rendering redundant devices costly and energy-intensive to communication systems. 
To pursue the goal of GIoT, it is imperative to reduce the number of transceivers required.

\begin{figure}[t]
	\centering
	\includegraphics[width=0.9\linewidth]{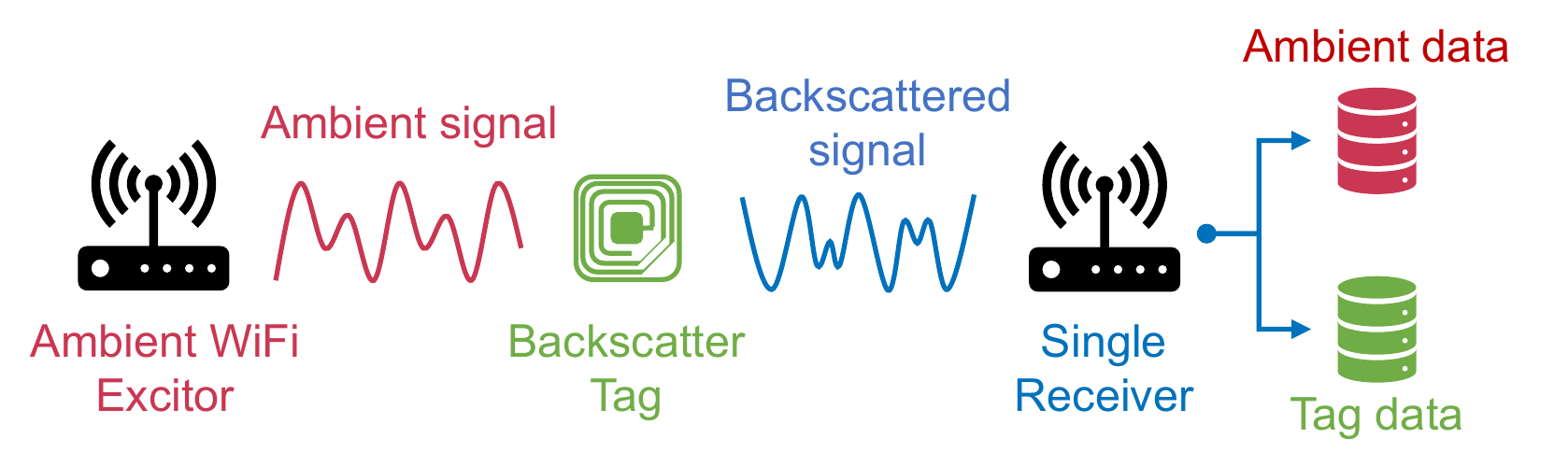}
	\caption{The framework of EEWScatter.}
	\label{framework}
\end{figure}

\subsection{EEWScatter Solution}

Although backscatter represents an attractive option for achieving low-power communication, existing systems do not fully utilize available resources. 
A significant challenge in developing an energy-efficient backscatter system is how to demodulate both ambient data and tag data using a single receiver, with the ambient signal serving as the excitation. 
In this paper, we present EEWScatter, a novel WiFi backscatter system that addresses this bottleneck.
Our approach leverages the reversible cyclic redundancy check (CRC) algorithm used in WiFi's frame check sequence (FCS) to recover ambient data. 
With this insight, we develop a CRC-based demodulation method that enables the use of ambient signals as carriers and a single receiver to demodulate both ambient and tag data. 
This approach reduces deployment costs and avoids interfering with original communication. 
Our experimental results demonstrate that EEWScatter achieves simultaneous transmission of both types of data with minimal energy consumption.

\section{EEWScatter Design}

The framework of EEWScatter is shown in Fig.\ref{framework}. 
In our proposed system, we use ambient WiFi signals as carriers for tag modulation. The tag then loads its data onto the carriers and backscatters it to the receiver. We present a novel demodulation method called CRC-Reverse to enable the demodulation of both types of data by a single receiver. 
%Further details will be provided in the following sections.
%The basic idea of them is to try to get one of the tag data and ambient data first, and then decode the other one.
%CRC-Verification generates a tag candidate set first and then recovers ambient data by the checksum function of CRC to find the correct data pair.
%Taking advantage of the reversibility of CRC Algorithm, CRC-Reverse first recovers the ambient data and then obtains tag data. 

\begin{figure}[t]
	\centering
	\includegraphics[width=0.75\linewidth]{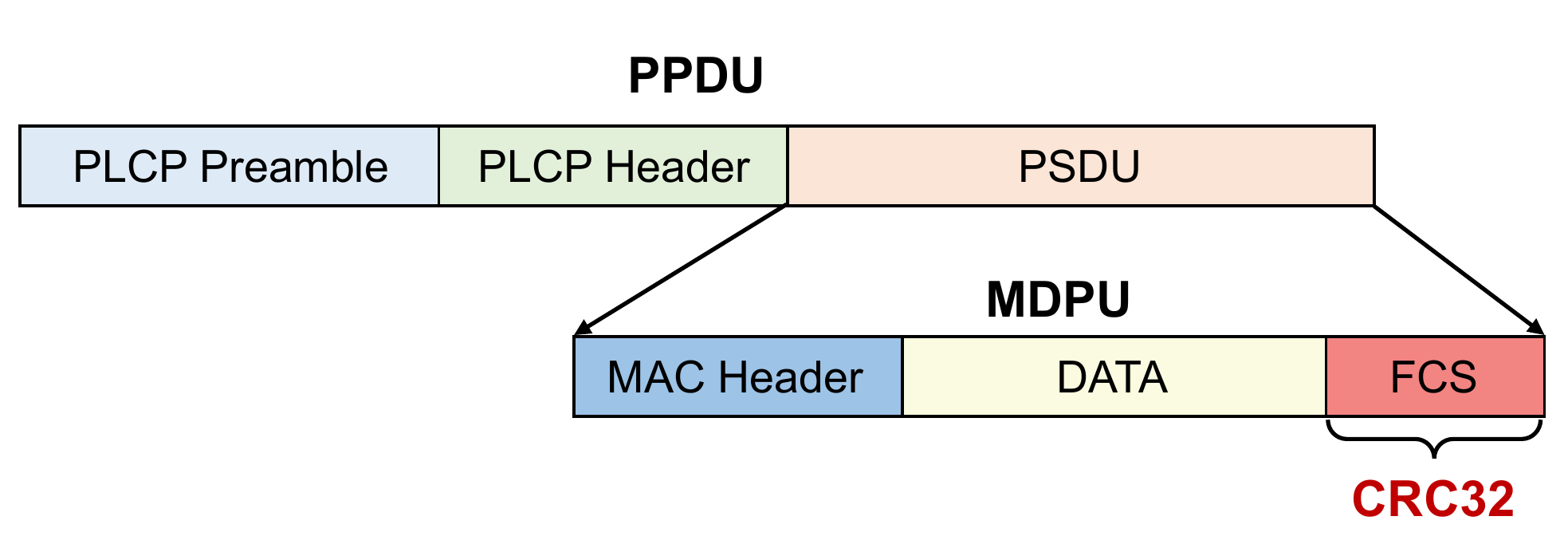}
	\caption{The structure of PHY frame and MAC frame of OFDM WiFi.}
	\label{OFDM WiFi}
\end{figure} 

\subsection{OFDM WiFi Excitation}
In this paper, instead of relying on an extra carrier generator, we reuse ambient signals as the excitation to avoid unnecessary excitor energy consumption. 
To achieve this, we need to select an appropriate excitation signal for backscatter first. 
There are several commercial signals available, including Bluetooth, ZigBee, LoRa, and WiFi. 
However, Bluetooth's short communication range and high mobility make it an unstable choice. 
Similarly, the limited deployment of ZigBee and LoRa devices incurs additional costs for deploying new equipment. 
In contrast, WiFi offers wide and stable coverage in our daily life making it the best choice. 
Among various WiFi protocols, we choose OFDM WiFi (such as 802.11g/n/ac) due to its high transmission rate, which means that less power is consumed to transmit one unit bit.

Fig.~\ref{OFDM WiFi} illustrates the PHY and MAC frame structure of OFDM WiFi \cite{9502043}.
The PHY protocol data unit (PPDU) consists of PLCP Preamble, PLCP Header, and PSDU.
Among them, PSDU includes a MAC protocol data unit (MDPU), which is composed of the MAC Header, Data, and FCS.
In OFDM WiFi, CRC32 is selected for FCS.
Specifically, it uses a polynomial of order 32 to compute the checksum value of the MAC frame at the transmitter and detects errors in the packet at the receiver. 
Our work leverages the advantages of CRC algorithm, which will be introduced in detail in section \ref{CRC-reverse}.
\begin{comment}
\begin{figure}[t]
	\centering
	\includegraphics[width=\linewidth]{crc_verification.pdf}
	\caption{The scheme of CRC-Verification demodulation.}
	\label{CRC_Verification}
\end{figure} 
\end{comment}
\subsection{Phase Modulation on Tag}
We employ backscatter modulation to transmit tag data to reduce transmission power consumption. 
Generally, three common methods can be used for tag modulation: ASK, PSK, and FSK.
However, the tag cannot modify the amplitude or frequency of OFDM WiFi signal because such modification creates an invalid codeword in the backscattered signal\cite{zhang2017freerider}.
Thus, we use the phase information to convey tag data. 
We do phase rotation by controlling the RF-switch on and off, which consumes only a few hundred microwatts of energy.

We define the original signal as $A\cdot e^{j(2\pi ft+\phi)}$, where $A$ denotes the amplitude and $2\pi ft+\phi$ represents the phase.
Phase modulation multiplies a phase change $\Delta\phi$ to the original signal.
Then the backscattered data can be expressed as $A\cdot e^{j(2\pi ft+\phi+\Delta\phi)}$. 
In our work, BPSK is utilized for tag modulation, and we apply different phase rotations to the original data to modulate different data bits.
As shown in Table \ref{phase modulation}, it rotates $0^\circ$ ($\Delta\phi =0$) if the tag data is bit 0, and rotates $180^\circ$ ($\Delta\phi=180^\circ$) if the tag data is bit 1.
The codeword translation \cite{zhang2016hitchhike} reveals that we can recover the tag data ($T$) from the original data ($O$) and the backscatter data ($B$) by $\quad B \oplus O = T$.
Note that to avoid corrupting other parts of the packet, we modulate only the data part of the PSDU and embed 1 tag bit on a single symbol.
%For example, if the received backscatter codeword is 1 and the corresponding ambient-codeword is 0, the tag-codeword is demodulated to 1.

\begin{comment}
\begin{equation}
    O \oplus T = B; \quad B \oplus T = O; \quad B \oplus O = T; 
    \label{codeword transalation}
\end{equation}
\end{comment}

%\begin{comment}
\begin{table} [t]
    \centering 
    \caption{Tag modulation based on phase rotation}
    \label{phase modulation}
    \begin{tabular}{|c|c|c|c|c|}
         \hline
         {Original data} & Tag data & {Phase rotation}& {Backscatter data}  \\
         \hline
         0 & 0 & $0^ \circ$ & 0  \\
         \hline
         0 & 1 & $180^ \circ$ & 1 \\
         \hline
         1 & 0 & $0^ \circ$ & 1  \\
         \hline
         1 & 1 & $180^ \circ$ & 0  \\
         \hline
    \end{tabular}
\end{table}
%\end{comment}

%Inspired by \eqref{codeword transalation}, we have an interesting discovery. 
%Except for the first formula used for tag modulation, the last two formulas can be used for data demodulation in the next section.
%Based on that, we design two new demodulation methods enabling a single receiver to recover both ambient data and tag data, which will be introduced in the next section.

\subsection{CRC-Reverse Demodulation} \label{CRC-reverse}

%In this section, we take full advantage of CRC and design a demodulation method: CRC-Reverse. 
%PLEASE explain it respectively
%They first recover the tag data and ambient data, respectively, and then use the last two formulas in \eqref{codeword transalation} to recover the other data.
To minimize overall power consumption, it is crucial to eliminate equipment redundancy.
The key problem is how to demodulate both ambient data and tag data from ambient backscattered signals with a single receiver. A possible solution could be to first demodulate the ambient data and then perform an XOR operation between the demodulated ambient data and backscattered data to recover tag data. However, the challenge lies in obtaining a reference for demodulating the ambient data. In other words, we need to identify an invariant that is related to ambient data and remains unaffected by tag modulation. Fortunately, FCS offered by the MAC layer provides the key to such demodulation. FCS, which encompasses the CRC for both the MAC Header and Frame Body, serves as a reliable reference. Importantly, even after backscattering, the FCS field of the packet remains unchanged, facilitating the demodulation process. 

However, there are still two main challenges to achieve successful data demodulation.

{\it 1) How to reverse the input data by CRC algorithm?}

First, it is essential to grasp the principles underlying the CRC algorithm. 
CRC is a short fixed-length datum (checksum) for an arbitrary data block.
Specifically, the algorithm works by performing a modulo-2 division of the input binary data D by a fixed binary number $A$, which is the coefficients of the generator polynomial $g(x) = \sum_{k = 0}^{r} a_i x^k$ with an order of R.
The remainder obtained from this division serves as the checksum, with a length of R. From a mathematical perspective, modulo-2 division can be represented as matrix operations. In particular, a generator matrix $G$ is derived by the generator polynomial $g(x)$ according to the length of input data \cite{generator}. If the input data $D$ is represented as a row vector of size 1*n, $G$ will have a size of n*R. The rank of $G$ does not exceed R. The CRC of $D$ can be expressed as \eqref{CRC computation}.
\begin{equation}
    C = D * G 
    \label{CRC computation}
\end{equation}

\begin{comment}
\begin{figure}[t]
	\centering
	\includegraphics[width=\linewidth]{crc.pdf}
	\caption{The scheme of CRC algorithm and CRC Reverse algorithm.}
	\label{CRC}
\end{figure}
\end{comment}

\begin{figure}[t]
	\centering
	\includegraphics[width=0.8\linewidth]{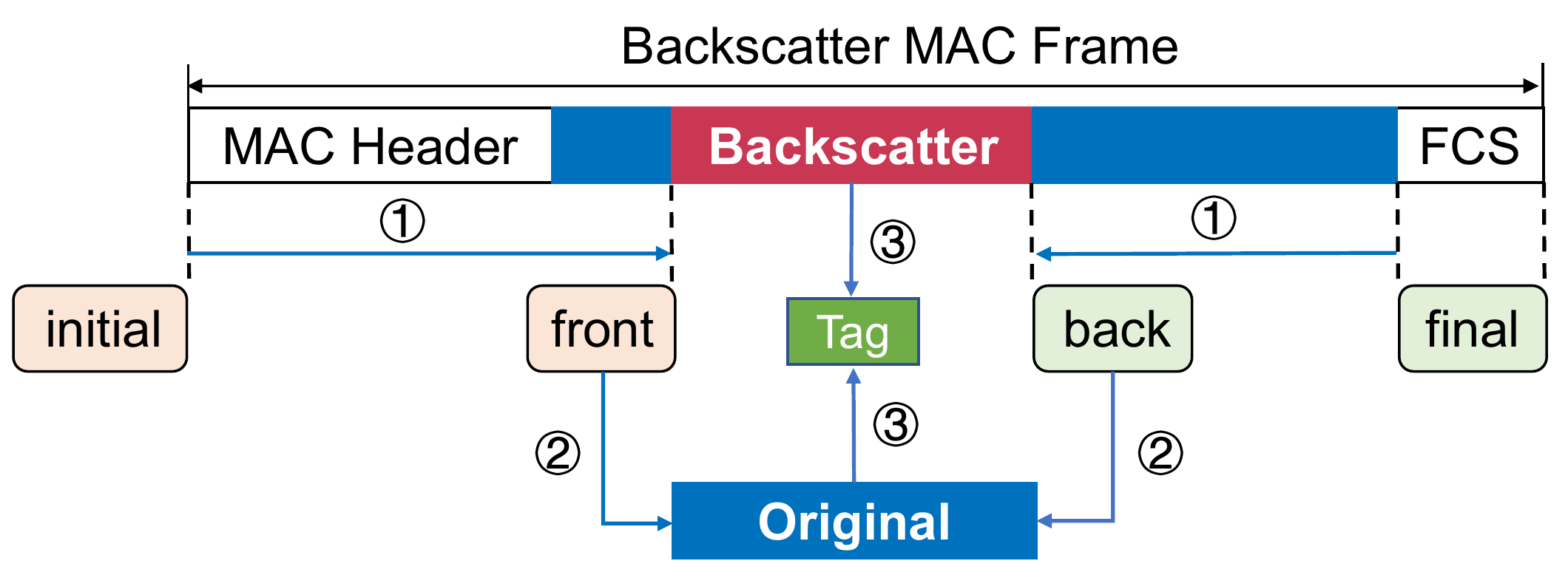}
	\caption{The scheme of CRC-Reverse demodulation.}
	\label{CRC_Reverse}
\end{figure}

Then we consider the hardware implementation. 
A shift register is used for checksum storage and CRC cyclic calculation. This register is typically initialized with a specific state value $I$. 
We can obtain the final state of CRC register $S$ by \eqref{CRC Generator}. 
\begin{equation}
    S = T^n(I) \oplus C 
    \label{CRC Generator}
\end{equation}

In this formula, $T^n(I)$ denotes the operation where the initial value of the register undergoes n iterations of Left Shift-XOR operation from the initial all-0 data sequence. In contrast, $T^{-n}(I)$ is the inverse of $T^n(I)$ and is defined as performing n iterations of XOR-Right Shift operation on $I$. Note that $T(\cdot)$ and $T^{-1}(\cdot)$ depend only on the input data length n and initial register $I$, bearing no relation to the data content. 
Since the size of $G$ is n*R, the CRC generation process is reversible when the input data length n is equal to R, and we can reverse the input data by \eqref{CRC computation} and \eqref{CRC Generator}. Specifically, if the initial value $I$ and final value $S$ of the register are known, we can obtain the checksum $C$ through a simple XOR operation by \eqref{CRC Generator}. 
Subsequently, we can perform matrix calculations to derive the input data as $D = C * G^{-1}$ by \eqref{CRC computation}.
Therefore, we can obtain a unique solution when n = R according to the properties of matrices.

{\it 2) How to recover both ambient data and tag data from backscattered OFDM WiFi signals?}

We have proved that a certain length of input data can be reversed by the initial register $r_{initial}$ and final register $r_{final}$.
Based on that, we propose a CRC-Reverse demodulation algorithm as shown in Fig.~\ref{CRC_Reverse}. 

The first step is to get the front register $r_{front}$ and back register $r_{back}$ of the backscattered data $B$.
We denote the data sequence before $B$ as $D_1$, the data length as $n_1$, and the generator matrix as $G_{n_1}$ 
and the corresponding data parameters after $B$ as $D_2$, $n_2$, and $G_{n_2}$.
Then $r_{front} = T^{n_1}(r_{initial}) \oplus (D_1 * G_{n_1})$ and $r_{back} = T^{-n_2}((r_{final}) \oplus (D_2 * G_{n_2}))$.

Subsequently, we use the two registers, $r_{front}$ and $r_{back}$, to recover ambient data.
In the MAC layer of 802.11n, the CRC32 is employed for FCS, where the generator polynomial $g(x)$ has an order of 32.
Therefore, to obtain a unique and definite solution, we can demodulate a data sequence of 32 bits.
In this scenario, the data length n is 32, the initial state $I$ in \eqref{CRC Generator} is $r_{front}$, and the final state $S$ in \eqref{CRC Generator} is $r_{back}$.
Then the ambient data $D_{32}$ can be recovered by \eqref{CRC Reverse}.

\begin{equation}
    D_{32} = (r_{back} \oplus T^{32}(r_{front}) )* G_{32} ^{-1}. 
    \label{CRC Reverse}
\end{equation}

Now, both the ambient data sequence ($O_{32}$) and backscattered data sequence ($B_{32}$) are known.
The final step is to recover the tag data.
However, the reversed ambient data is in the MAC frame, while the tag phase modulation conducts on PHY data.
The data mapping relationship between the MAC and PHY layers should be considered.
When the modulation coding scheme (MCS) is set to 0 and the channel bandwidth is set to 20MHz, a single symbol in PHY of 802.11n contains 26-bit MAC data. Since our tag modulates a bit on one symbol and $\lfloor \frac{32}{26} \rfloor = 1$, we can recover enough original data for further tag demodulation. 
Now, the tag data sequence $T_{32}$ can easily be demodulated by XOR: $T_{32} = O_{32} \oplus B_{32}$.
Finally, we translate the tag bit as bit '1' when over half of the bits in $T_{32}$ is 1.
Otherwise, we translate it as bit '0'.

Although we have recovered both data successfully, there is room for further exploration.
In general, we can only modulate a single bit for one packet since there is only one CRC in a single packet.
To expand the modulating capacity, we propose a multi-CRCs mode where a packet can contain multiple CRCs.
The aggregate MAC protocol data unit (A-MPDU) makes this idea available.
Multiple MPDUs are aggregated into a single packet.
Each MPDU has a standard MAC frame containing an FCS.
A-MPDU is available in most WiFi routers.
Therefore, this mode does not entail any additional device overhead.

%\begin{figure}[htbp]
%\centerline{\includegraphics{fig1.png}}
%\caption{Example of a figure caption.}
%\label{fig}
%\end{figure}

\begin{table} [t]
    \centering 
    \caption{Overall-power-consumption Comparison}
    \label{overall comparison}
    \resizebox{0.95\linewidth}{!}{
    \begin{tabular}{|c|c|c|c|c|}
         \hline
         {Category} & Excitor & {Tag}& {Receiver} & Total \\
         \hline
         Passive WiFi* & 10W & 14.5$\mu$W & 5.4W & 
         $\sim$15.4W\\
         \hline
         FreeRider* & 0 & 33$\mu$W & 10.8W & $\sim$10.8W\\
         \hline
         EEWScatter & 0 & 271$\mu$W & 5.4W & $\sim$5.4W \\
         \hline
    \end{tabular}
    }
\end{table}

\begin{comment}
\begin{figure}[t]
    \centering
\begin{minipage}{0.45\linewidth}
		\centering
		\includegraphics[width=\linewidth]{time.pdf}
		\caption{Decoding time}
		\label{fig:1}
	\end{minipage}
	%\qquad
	\hfill
	\begin{minipage}{0.45\linewidth}
		\centering
       \tabcaption{Radio comparison}
        \resizebox{\linewidth}{!}{
        \begin{tabular}{|c|c|}
        \hline
        Radio & Power Consumption \\ \hline
        Active Radio & 108mW \\ \hline 
        EEWScatter Tag & 271$\mu$W \\
        \hline 
        \end{tabular}}

        \label{fig:NormalSafe}
\end{minipage}
\end{figure}
\end{comment}

\begin{figure}[t]
\centerline{
\subfloat[{Tag consumption comparison.}]
{\includegraphics[width=0.45\linewidth]{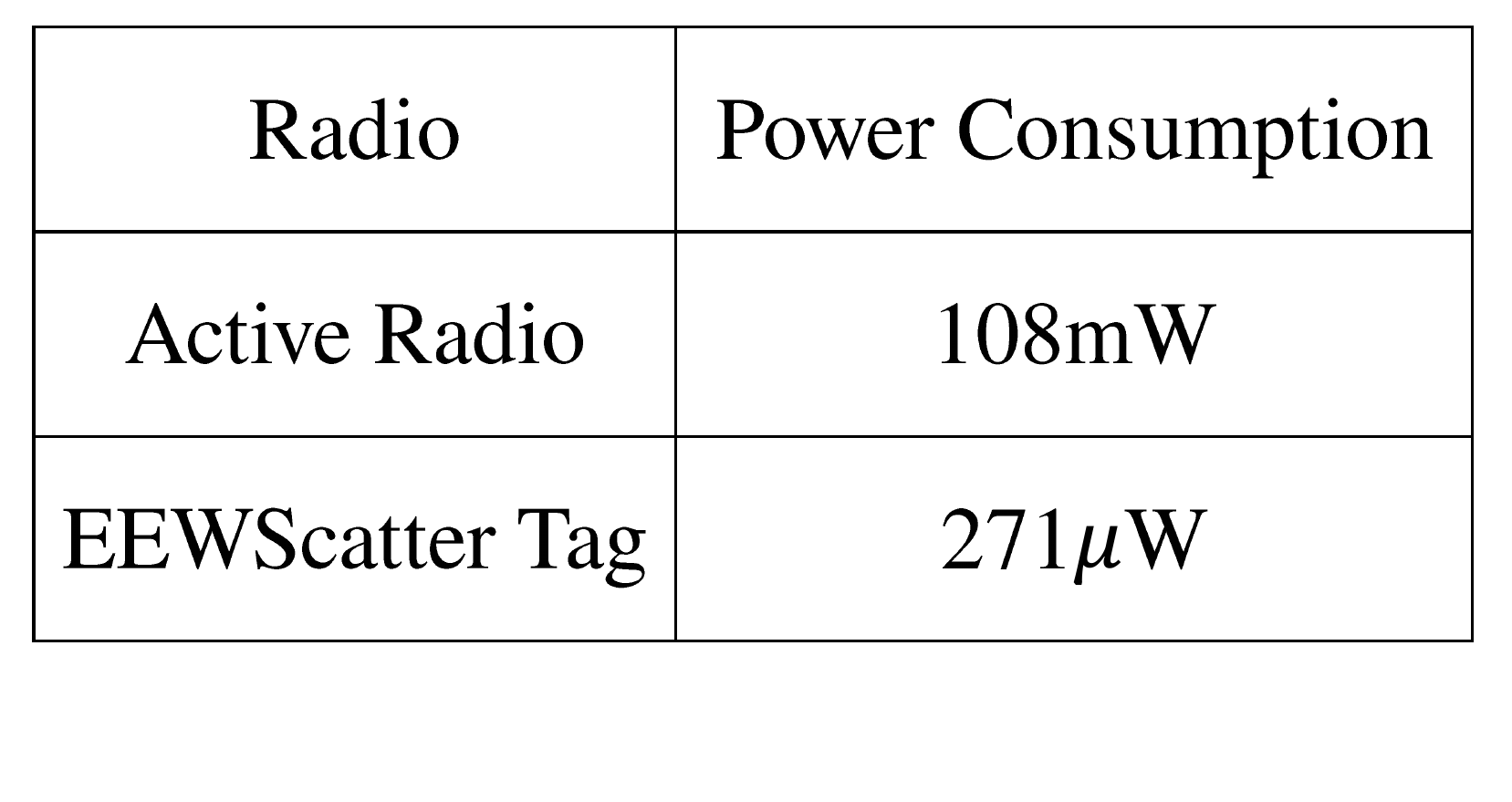}
\label{tag consumption}}
\hfil
\subfloat[{Decoding time.}]
{\includegraphics[width=0.5\linewidth]{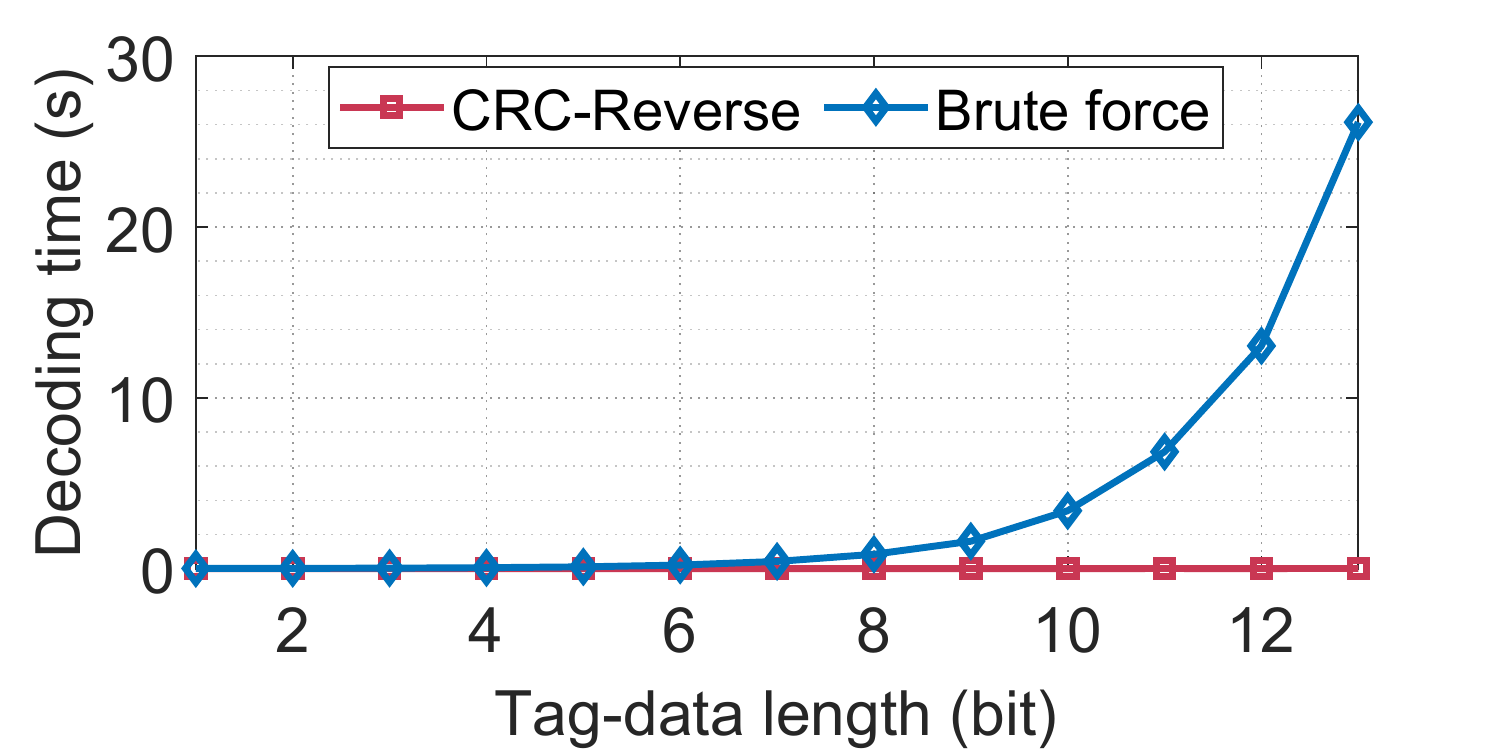}
\label{time}}
}
\caption{Tag power consumption and demodulation computational complexity.}
\label{translation}
\end{figure}

\section{Evaluation}

\subsection{Power Consumption}

First, we present a power consumption comparison of the entire system. 
Subsequently, we evaluate the individual performance of each component.
Section \ref{redundant transceiver} outlines that current backscatter systems can be categorized into two types: those that require an additional transmitter helper and those that rely on two receivers. 
Therefore, we select these two types as our baselines and represent them with Passive WiFi* and FreeRider*, respectively.

\textbf{Overall.}
To get close to the real-world situation, we use the power of commercial devices to evaluate the transceivers.
Specifically, we use the AnyID R342N RFID reader with a rated power of 10 W to represent the transmit helper and use the TP-LINK TL-WR842N commercial router with a rated power of 5.4 W to present the WiFi receiver.
Table \ref{overall comparison} illustrates the comparison between EEWScatter and two typical systems.
We can see that Passive WiFi* brings an additional consumption of 10 W since it requires a helper for carrier generation.
For FreeRider*, the redundant receiver creates an extra burden of 5.4 W.
In contrast, EEWScatter is excited by ambient traffic and decodes both data with only one receiver, eliminating transceiver redundancy significantly. 
Thus, EEScatter's system power consumption is about 1/2 that of FreeRider* and 1/3 that of Passive WiFi*.

\begin{figure}[t]
\centerline{
\subfloat[{Comparison for excitor.}] {\includegraphics[width=0.5\linewidth]{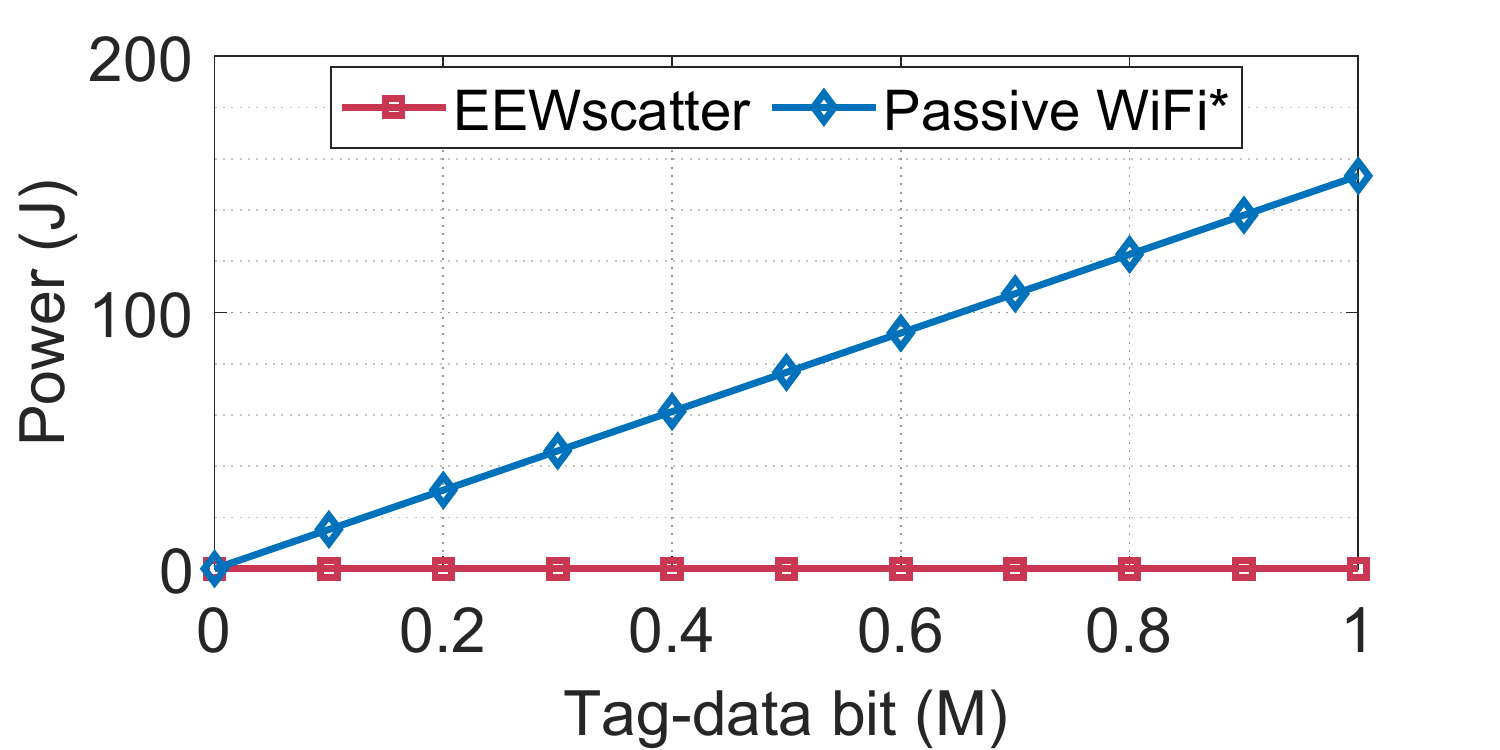}
\label{power excitor}}
\hfil
\subfloat[{Comparison for receiver.}]
{\includegraphics[width=0.5\linewidth]{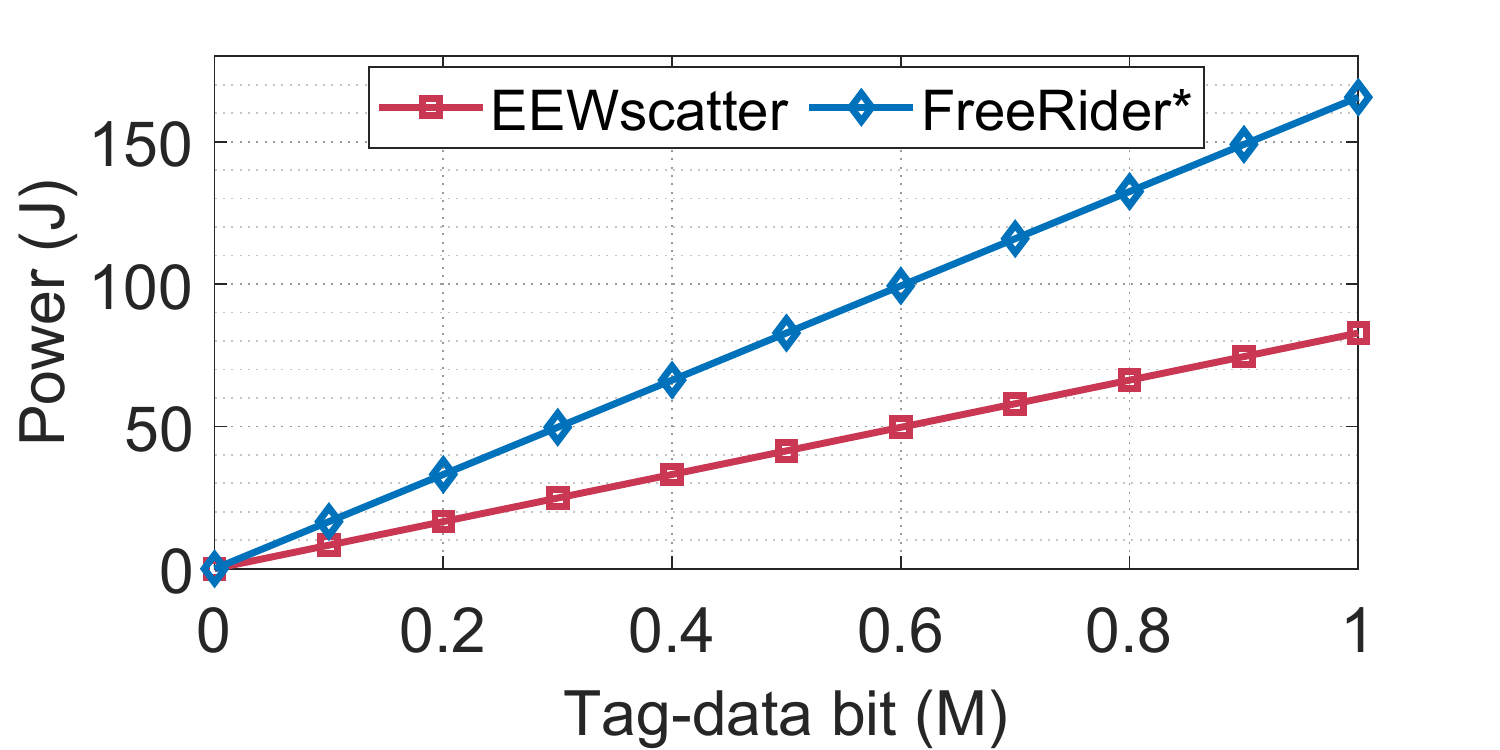}
\label{power receiver}}
}
\caption{Comparison of power consumption for transceivers.}
\label{Power consumption}
\end{figure}

\begin{figure}[t]
\centerline{
\subfloat[{Packet recovery rate.}] {\includegraphics[width=0.5\linewidth]{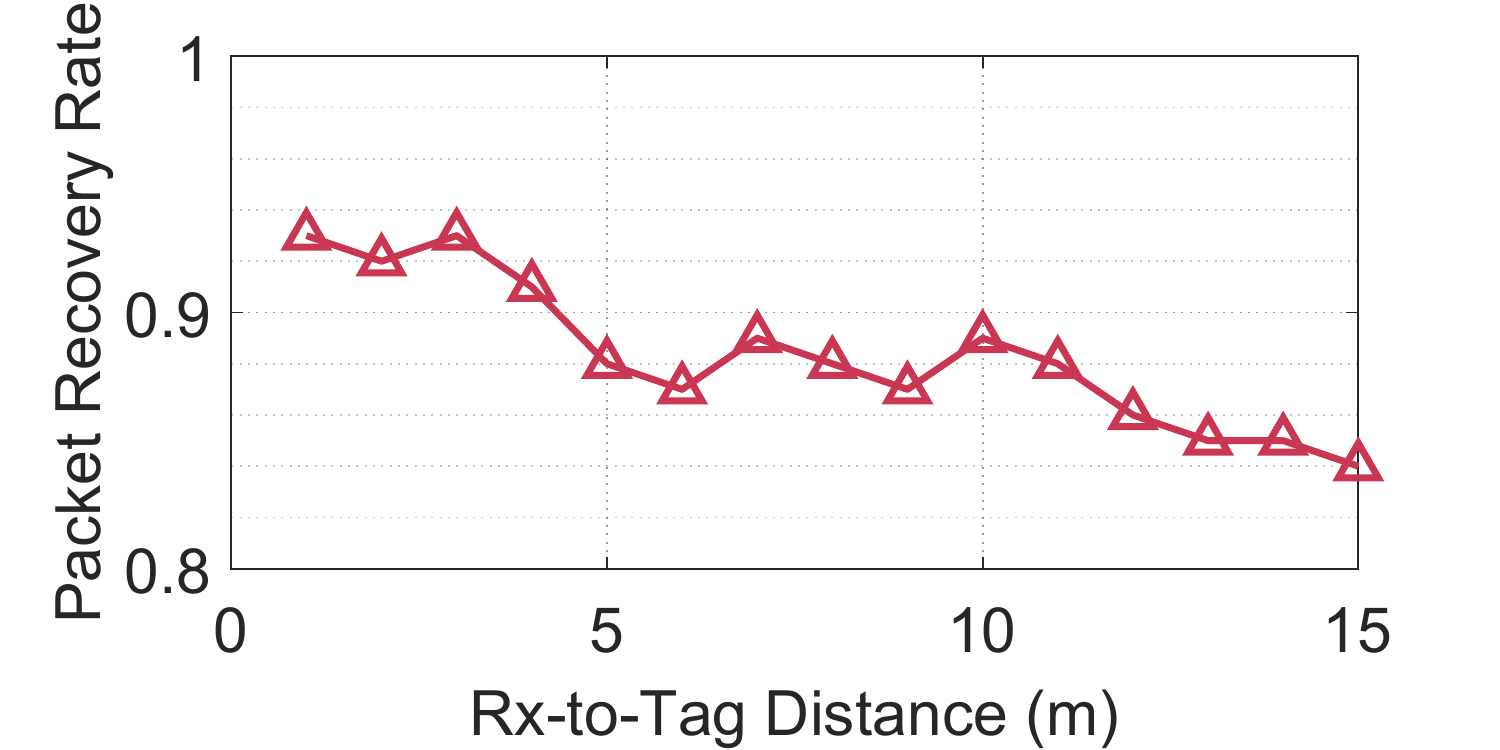}
\label{packetrecoveryrate}}
\hfil
\subfloat[{BER of tag data.}]
{\includegraphics[width=0.5\linewidth]{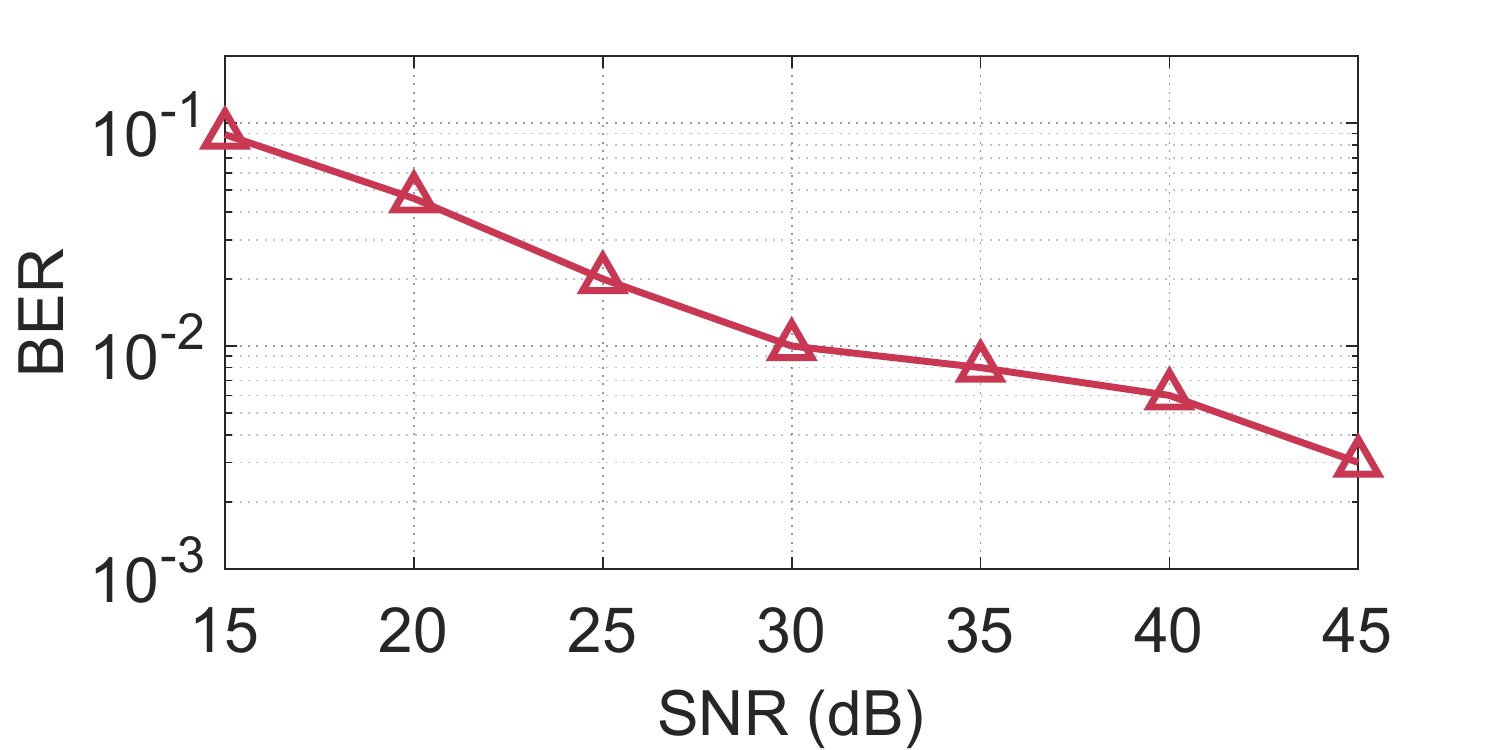}
\label{ber}}
}
\caption{Transmission performance of CRC-Reverse.}
\label{transmission performance}
\end{figure}

\textbf{Breakdown.}
Next, we evaluate the power consumption comparison for each component and the computational complexity of our demodulation algorithm. 

1) Tag comparison: 
The power consumption of the tag obtained by IC simulation is approximately 271 $\mu$W.
In comparison, Broadcom BCM4330, a chip supporting active WiFi radio, consumes 108 mW in active mode, which is about 1000x of EEWScatter tag.
It seems that the power consumption of our tag is slightly higher than the other two systems.
That is because we perform signal synchronization on the tag to precisely align the position of the tag modulation.
These operations effectively improve system performance with only a slight increase in power consumption.

2) Transceivers comparison:
Our study focuses on quantifying the power consumed by transceivers for transmitting a specific amount of tag data.
The power consumption is calculated by multiplying transmission time by the rated power.
%We use the same method to measure the transceivers and evaluate them separately.
As shown in Fig.~\ref{power excitor}, the power consumption of Passive WiFi* increases rapidly as more data is transferred due to the use of help device. 
In comparison, there is almost no consumption for our excitor since we use the ambient signal as excitation.
Fig.~\ref{power receiver} demonstrates that the redundant receiver for FreeRider* results in at least 2x overhead of that of our system.
%The synchronization between two receivers further increases system power consumption.
As data transmission volume increases, the gap in power consumption becomes more significant.

3) Computational complexity:
We measure the computational complexity in terms of decoding time and compare our approach to brute force search, which is a naive solution to this problem.
Brute force search first identifies all possible tag data candidates, then recovers the original data, and finally utilizes CRC detection to verify the intended target.
It's evident that brute force research has a complexity of O($2^n$), whereas our CRC-Reverse algorithm has a complexity of O(n).
The results depicted in Fig.~\ref{time} are consistent with our theoretical analysis.
Thus, our system incurs less computational consumption.

\begin{comment}
\subsection{Application Case: Greenhouse} 
Green IoT not only involves Green in IoT, but also Green by IoT.
Greenhouse provides a suitable growing environment for crops by intelligently controlling the temperature, humidity, etc.
%Therefore, a large number of sensors are needed.
%Active devices are not suitable for this case as the cost of regular device maintenance and battery replacement is high and the batteries will contaminate the soil affecting crop growth.
%EEWscatter provides an energy-efficient and environment-friendly solution for environmental sensing.
To collect environmental information, we apply our battery-free tags in the soil or over the air to measure and transmit data like temperature, soil moisture, pH, and so on.
WiFi routers placed in the greenhouse provide excitation signals for tags. 
These data are collected and uploaded to the system control center for intelligent decision-making.
Then the system can choose the optimal strategy to consume minimal resources such as water, fertilizer, and air conditioning to keep a suitable greenhouse environment.
The farmer even can use a single mobile phone to monitor the greenhouse conveniently.
\end{comment}

\begin{comment} 
\begin{figure}
	\centering
	\includegraphics[width=0.8\linewidth]{greenhouse.pdf}
	\caption{Greenhouse.}
	\label{greenhouse}
\end{figure}
\end{comment}

\begin{figure}[t]
\centerline{
\subfloat[{Backscatter system implementation.}] {\includegraphics[width=0.5\linewidth]{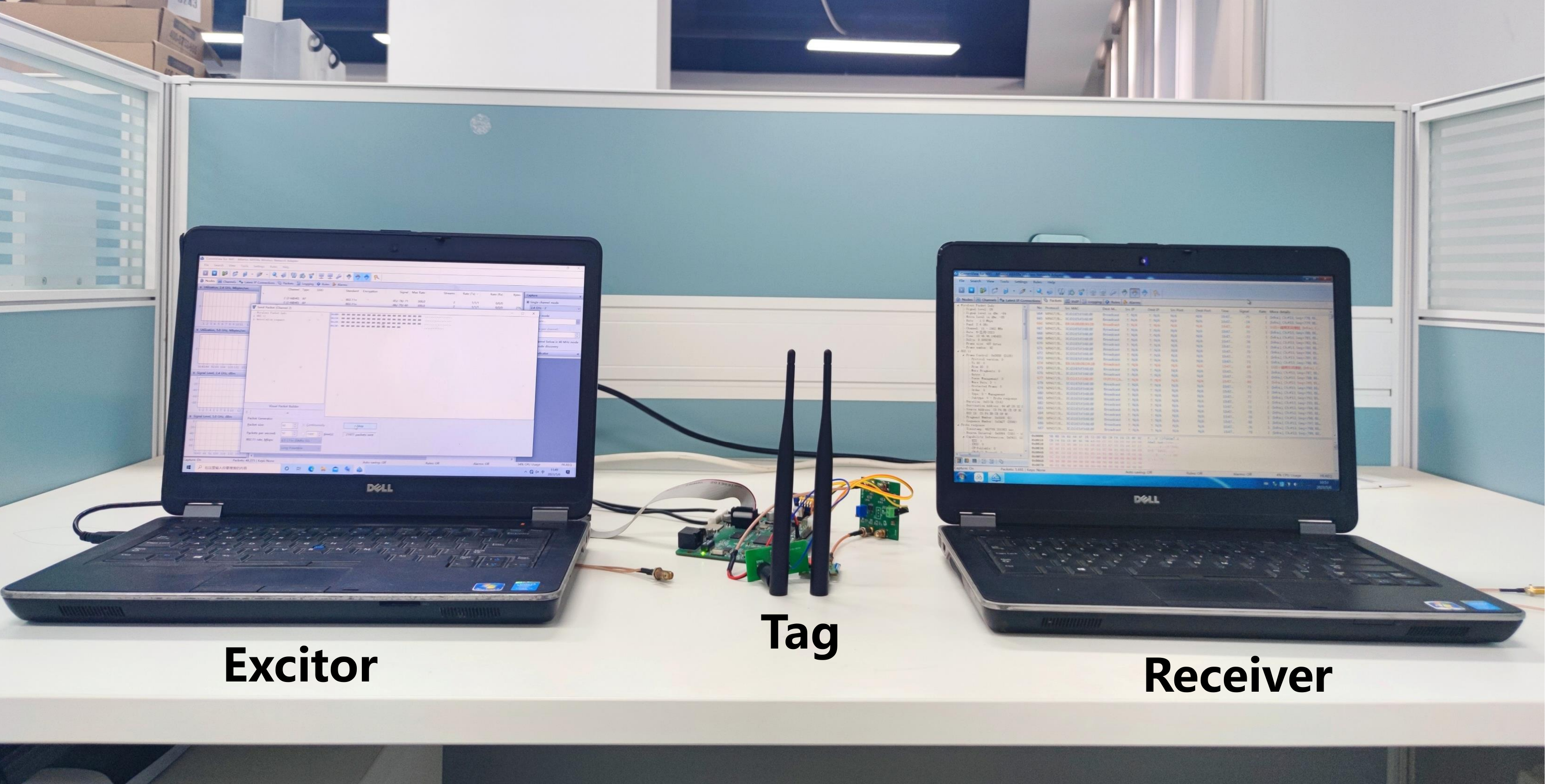}
\label{device}}
\hfil
\subfloat[{RSSI of backscattered data.}]
{\includegraphics[width=0.5\linewidth]{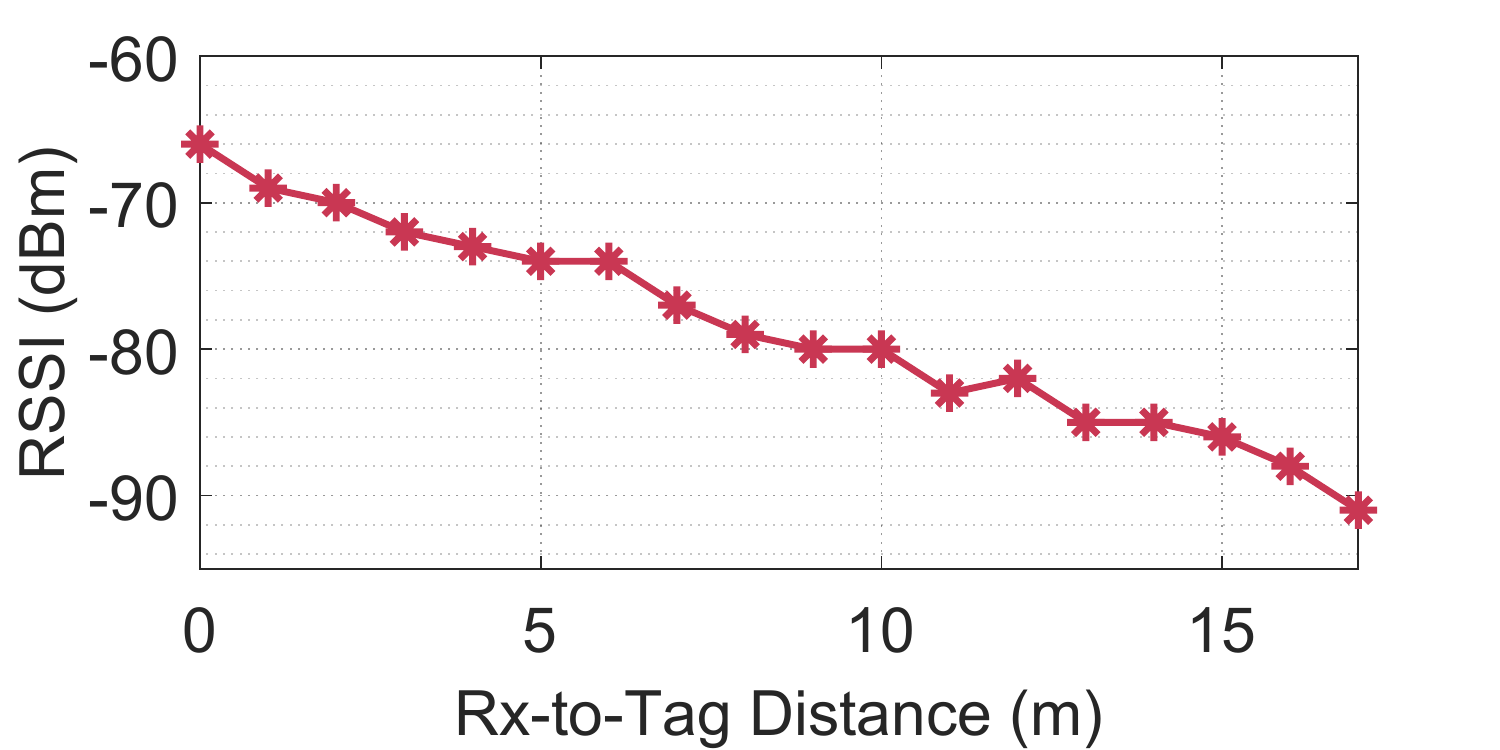}
\label{RSSI}}
}
\caption{Hardware implementation.}
\label{hardware}
\end{figure}

\subsection{Transmission Performance}
\textbf{Simulation Setup.}
We evaluate the transmission performance of EEWScatter by simulating it with MATLAB.
Specifically, we use 802.11n signals with MCS=0 as excitations and use BPSK to do tag modulation.
To replicate realistic channel conditions, we set the delay profile of the 802.11n multipath fading channel to Model-B.
In our experiments, we disable BCC (binary convolutional code) and use A-MPDU frame structure conveying 10 tag bits per packet.

\textbf{Simulation Results.}
We evaluate the packet recovery rate (PRR) and the tag bit error rate (BER) in this section.
1) Packet recovery rate: 
We define the ratio of correctly decoded ambient data as packet recovery rate, which will directly affect the accuracy of tag data decoding.
We vary the distance between the receiver and the tag from 0 to 15 m and find that PRR decreases slightly with increasing distance as shown in Fig.~\ref{packetrecoveryrate}. 
However, it still remains above 90\% within 5 m.
2) Tag-data BER: 
In general, a low BER indicates the accuracy of data transmission and the reliability of the system.
When measuring BER, we fix the distance between the tag and the receiver to 10 meters. 
Fig.~\ref{ber} shows that BER is below $10^{-2}$ when SNR$>$30, which meets the requirements for reliable communication.

\textbf{Hardware setup and results.}
We prototype the tag using a signal detector constructed
by an AD8313 and TLV3501, a Xilinx Artix-7 FPGA and an RF switch.
The WiFi transceivers are prototyped using a Dell laptop equipped with a Qualcomm Atheros AR938x. 
We use commercial software CommView to facilitate the control of ambient WiFi signal and measure the RSSI of received signal.
We send the 802.11n with 6.5 Mbit/s at channel 2, and capture the backscattered data at channel 12.
Fig.~\ref{RSSI} shows that the RSSI of backscattered signal is over -80 dBm within 10 m providing assurance for further data demodulation, which will be done in our future work.

\section{Discussion and Future Work}
\subsection{BCC-enabled Solution}
We turn BCC off as the error-correcting ability makes the backscatter data different from our expectation in this system.
However, we acknowledge that BCC is useful for OFDM WiFi, and we have identified some specific encoding patterns in BCC that can facilitate demodulation. 
Hence, we will design a BCC-enabled system in our future work.

\subsection{High-order Excitation}
%In this paper, we use BPSK for tag modulation.
%In the future, we aim to explore higher-order modulation like QPSK and QAM to increase the throughput.
%That can decrease the energy consumed to transmit a single bit and enable our system to be applied to more scenarios.
%CRC-Reverse has a restriction of MCS=0 on ambient OFDM-WiFi signals.
In this paper, we impose an MCS = 0 constraint since the MAC data contained by a PHY symbol exceeds 32 bits in higher-order excitations making data demodulation more challenging. 
%For higher-order ambient signals, a single symbol exceeds 32 bits, which increases the difficulty of demodulation.
%We have found that CRC-Reverse has the potential to demodulate symbols exceeding 32 bits.
Our investigations indicate that CRC-Reverse may have the capability to demodulate the input data over 32 bits when some new constraints are added.
To enhance the universality of GIoT, we are going to optimize the CRC-Reverse demodulation algorithm later to handle high-order excitations.

\section{Conclusion}
In an era when IoT systems are consuming more resources and polluting the environment rapidly in our living space, EEWScatter provides a green solution to realize GIoT.
We are the first to consider power consumption from a system-wide perspective and propose a scheme enables all three main components of the system to be energy-efficient.
In particular, we have designed a new CRC-based demodulation method to eliminate the redundancy of the additional transceiver commonly used in the current systems.
Our approach saves at least half of the power consumption of existing systems, and it can be used in intelligent control systems like greenhouse environmental control system to further reduce the waste of resources.
We hope our work can trigger further thinking about GIoT and fuel more researchers to work along this line.

%\section*{Acknowledgment}

%\bibliographystyle{IEEEtran}
%\bibliography{reference}
% Generated by IEEEtran.bst, version: 1.14 (2015/08/26)

\end{document}